\documentclass[12pt]{article}
\usepackage{epsfig,graphics}
\usepackage{procrpc4,color}

\def\c2{CLEO~II.V}

\def\d0d0{ D^0\bar{D}^0 }
\def\p0p0{ P^0\bar{P}^0 }
\def\qp2{ \Bigl| \frac{q}{p} \Bigr|^2 }
\def\pq2{ \Bigl| \frac{p}{q} \Bigr|^2 }

\def\ee{ e$^+$e$^-$ }

\def\be{\begin{equation}}
\def\en{\end{equation}}
\def\bea{\begin{eqnarray}}
\def\eea{\end{eqnarray}}

\def \epem{e^+e^-}

\begin{document}


\begin{flushright}
FERMILAB-Pub-03/331-E\\
FRASCATI LNF-03/18(P)\\
\end{flushright}
\begin{center}
{\Huge\bf On the narrow dip structure at 1.9~GeV/c$^2$ 
in diffractive photoproduction}
\\
\vspace{.5cm}
Submitted to Phys. Lett. B
%
%
\begin{center}

\normalsize

\bigskip

P.L.~Frabetti$^a$,
H.W.K.~Cheung$^{b,1}$, 
J.P.~Cumalat$^b$, 
C.~Dallapiccola$^{b,2}$, 
J.F.~Ginkel$^{b}$               
W.E.~Johns$^{b,3}$,             
M.S.~Nehring$^{b,4}$,           
E.W.~Vaandering$^{b,3}$,                
J.N.~Butler$^c$, 
S.~Cihangir$^c$
% ...

\bigskip

\small

$^a$ {\it Dip.   di   Fisica   dell'Universit\`{a}   and   INFN-Bologna,  
I-40126   Bologna,   Italy.}

$^b$ {\it  University   of   Colorado,   Boulder,   CO   80309,  USA.}

$^c$ {\it  Fermi   National   Accelerator   Laboratory,   Batavia,  
  IL   60510,   USA.}

\end{center}
\begin{flushleft}
% ...
$^1$ Present address:  Fermi National Accelerator Laboratory, Batavia, IL
60510, USA. 

$^2$ Present address:  University of Massachusetts, Amherst, MA 01003, USA.

$^3$ Present address:  Vanderbilt University, Nashville, TN 37235, USA.

$^4$ Present address:  Adams State College, Alamosa, CO 81102, USA.

% ...
Science, Yusong P.O. Box 102, Taejon 305-600, South Korea. 

\end{flushleft}
\bigskip

$PACS:$ 13.25.Jx, 13.60.Le, 14.40.Cs

\bigskip

\today
%
%
\begin{abstract}
The narrow dip observed at 1.9~GeV/c$^2$ 
by the Fermilab experiment E687 in diffractive photoproduction of
$~3\pi^+3\pi^-$ is examined. 
The E687 data are refitted, a mechanism 
is   proposed  to explain why this resonance appears 
as a dip, and possible interpretations are discussed.
\end{abstract}
\end{center}
\section{Introduction.}
The E687 experiment at Fermilab has observed \cite{e687} a narrow dip at
M~ = ~1.911 $\pm$ 0.004 $\pm$ 0.001 ~GeV/c$^2$ and with a width $\Gamma$ = 29 $\pm$
11 $\pm$ 4 ~MeV/c$^2$ 
in $ 3\pi^+ 3\pi^- $ diffractive photoproduction.
 If interpreted as a resonance, it  has
J$^{PC}=1^{--}$ quantum numbers, G=+1 because of the six-pion final state and
consequently I=1. 
The structure found by E687 recalls what was observed with lower
statistical significance by the 
DM2 collaboration \cite{baldini0} \cite{Donnachie}, in the channels
$e^+ e^- \rightarrow 3\pi^{+}3\pi^{-}$ 
and $\epem \rightarrow 2\pi^{+}2\pi^{-}2\pi^{0}$.
BABAR is investigating the same channels by means of initial state
radiation with better statistical significance than DM2, thanks
to the very high integrated luminosity provided by PEP-II.
\par 
In this paper we refit
the E687 data and  discuss the extent to
which this new resonance interferes with known vector resonances.
We propose a mechanism, pointed out for somewhat similar circumstances
\cite{Gilman},  
to explain 
why this resonance appears 
as a dip and examine some possible interpretations.
\par
\section{Fitting procedure and resonance parameters}
 It is difficult to obtain spin and parity of a six-pion final state. 
Supported by the E687 diffraction photoproduction data, we assume that in the
 selected experimental conditions 
the incident photon energy is
high enough and the momentum transfer to the target small enough to
fulfill naive diffractive photoproduction expectations, namely:
\begin{enumerate}
\item photon quantum numbers are
transferred to the produced hadronic mass M;
\item Vector Meson Dominance \cite{VDM} holds, i.e., 
diffractive photoproduction cross section and \ee annihilation at a
c.m. energy M are related, for a given final state of mass M, as follows: 
\be
 \sigma_{\gamma N \rightarrow V N}^{diff} \propto \Gamma_{V}^{ee}  
 \cdot \sigma_{V N\rightarrow V N}
  \label{EQ:PHOTOEPEM}
\en
where
\be
 \Gamma_{V}^{ee} ~\sim~ \frac{1}{3\pi^2} \cdot \int dM~ \cdot M^2 
 \sigma_{e^+e^- \rightarrow V}( M)
  \label{EQ:SIGMAEE}
\en
\end{enumerate}
Consistent with this assumption, we expect the vector meson elastic cross
section $\sigma_{V N\rightarrow V N}$ 
 to vary very slowly as a function of M, depending  on the
V valence quark flavors. At the E687 photon-beam energies,
corrections due to the variation with M of the target form
factor (diffractive $t$-slope) are expected to be very small, since 
$t_{min} \sim \frac{M^4}{(2E_\gamma)^2} \sim 2\cdot 10^{-4} ~GeV^2$, to be compared to
the diffractive slopes $\sim 2\cdot 10^{-2} ~GeV^2$.
Provided the aforementioned assumptions are valid for a superposition of vector mesons
with the same valence quarks as in the case of an isovector final state,
by differentiating Eq.~\ref{EQ:PHOTOEPEM} and dropping V we expect the
following: 
\be
 \frac{1}{M^2} \cdot \frac{d\sigma_{diff}}{dM}_{\gamma N\rightarrow
 6\pi N}(M) \propto \sigma_{e^{+}e^{-} \rightarrow 6 \pi}(M). 
 \label{EQ:SIGMA}
\en
Therefore, the diffractive photoproduction mass
spectrum as a function of M, once
weighted by a factor $1/M^2$, can be directly compared to \ee
annihilation at the c.m. energy M.
The  fair agreement between
$e^{+}e^{-} \rightarrow 2 \pi^{+} 2 \pi^{-}$ \cite{babar4pi}
and the weighted diffractive
photoproduction of $2 \pi^{+} 2 \pi^{-}$ \cite{Lebrun:1997aa}
 supports this
relationship.
A better agreement would be obtained at high invariant masses assuming a 
mild dependence on M of the aforementioned factors.
In the following diffractive photoproduction data is
 considered weighted by the $1/M^2$ factor to facilitate comparison 
with \ee annihilation data.
\par
\begin{figure}[htb]
 \centerline{
   \epsfig{file=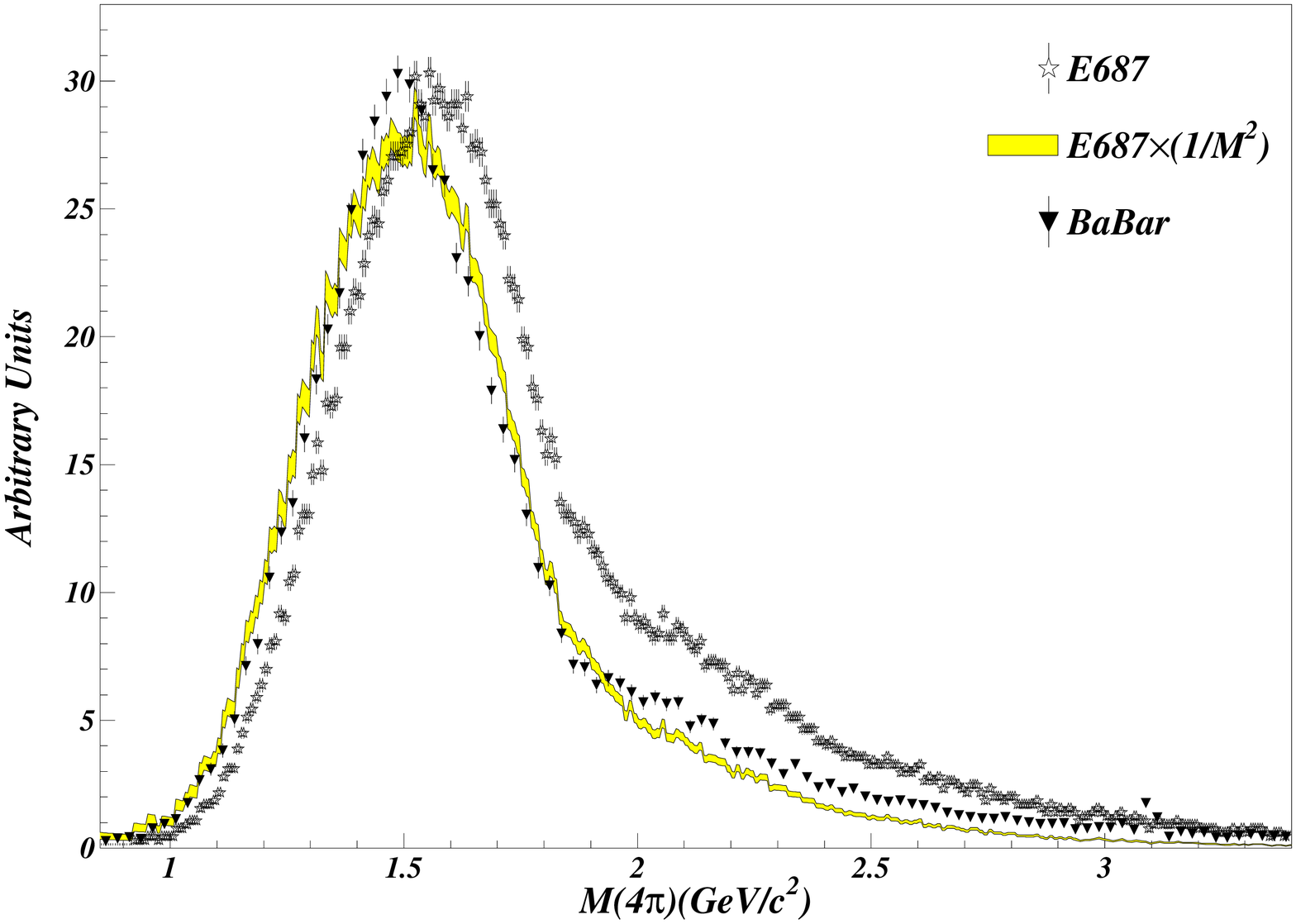,width=5.0in,height=5.0in}
 }
 \caption{BaBar  and E687  2$ \pi^{+} 2 \pi^{-}$ invariant mass
 distributions. The E687 acceptance 
 corrected yields (band) have been normalized to the BaBar cross-sections via
 Eq.~\ref{EQ:SIGMA}. }
 \label{FIG:BABARE687}
\end{figure}
\par
Data in \cite{e687} have been fitted by considering a narrow resonance $V_0$ and a
Jacob-Slansky continuum \cite{jacob}. In the Jacob-Slansky (J-S) model the diffractive
continuum 
is represented by an amalgamation of broad resonances, which may interfere with the narrow
resonance $V_0$:
\[
F_{JS}(M)=f_{JS}^2(M)=c_0+c_1\frac{e^{\frac{-\beta}{M-M_0}}}
{(M-M_0)^{2-\alpha}}.
\]
\par
In this paper we extract from the continuum another resonance
in addition to the $V_0$ narrow resonance.
We then perform a fit of the $1/M^2$ weighted data with two 
BW-resonances plus a function $f_{JS}(M)$, representing the background. 
This fit function, made of two resonances $V_{0,1}$ and
a background contribution $f_{JS}(M)$, describes the invariant mass distribution
in the whole accepted mass range $1.5\div 3.2 \;GeV$
 with $\chi^2/dof=1.06$, in the selected mass range $1.65\div 2.4 \;GeV$
with  $\chi^2/dof=0.80$ as shown in Fig. 2, 
and the resulting shape is similar to the one in \cite{e687}. 
The fit parameters are reported in Tab.~\ref{TAB:FIT1}.
The masses and widths of $V_{0,1}$ are consistent with the narrow resonance
in \cite{e687}, i.e., $M_0=~1.910\pm0.010~GeV$ and width $\Gamma_0=~37\pm 13~GeV$, 
and with the known vector recurrence $\rho(1700)$, quoted
in the PDB \cite{PDB}. Phases and partial widths are also reported in 
Tab.~\ref{TAB:FIT1}. 
Partial widths are given in arbitrary units and only their relative 
ratio is meaningful. This is due to the fact that the E687 data are presented 
as (efficiency corrected) yield, and not as a cross section.
\par
The function $F_{JS}(M)$ not only models the slowly rising continuum, which 
includes all the vector mesons resonances, but also any non-interfering incoherent 
background that might remain after statistically subtracting from the 
$3\pi^+ 3\pi^- $ invariant mass distribution. The level of the incoherent 
background is relatively high, about 30\%.  However, it is difficult to
estimate the magnitude of what remains after subtraction. According to the
relative phases there is a large interference of $F_{JS}$ with $V_{1}$,
which confirms that the residual incoherent background contribution is not important.  
\par
We also checked the effect of replacing the
$f_{JS}(M)$ amplitude with a broad $V_2$ Breit-Wigner amplitude, fit results
with three Breit-Wigner, shown in Tab.~\ref{TAB:FIT}, are consistent with the 
previous values of $V_0$ and $V_1$, in particular $M_0=~1.910\pm0.010~GeV$ 
and $\Gamma_0=~33\pm 13~GeV$.
 \begin{table}[h!]
   \begin{center}
     \begin{tabular} {|p{2.15cm}|p{2.99cm}|p{2.08cm}|c|c|} \hline
& \multicolumn{1}{|c|}{Mass} 
& \centering{Width} & $B_{ee}B_{3\pi^{+}3\pi^{-}}/M^2$ & Phase \\ 
 \vspace{-.0cm}
\centering{Resonances}   
& \centering{(GeV/c$^2$)} & \centering{(MeV/c$^2$)} & (Yield/10 MeV)& (deg.) \\ 
\hline
 \multicolumn{1}{|c|}{$V_{0}$} 
& \multicolumn{1}{|c|}{$1.910\pm 0.010$}
 & \centering{$37\pm 13$} & $ 5\pm 1$ & $10\pm 30$ \\
 \multicolumn{1}{|c|}{$V_{1}$} 
& \multicolumn{1}{|c|}{$1.730\pm 0.034$} 
& \centering{$315\pm 100$} & $ 17\pm 3$& $140 \pm 10$\\
 \hline
      \end{tabular}
     \vfill
   \end{center}
 \end{table}
\begin{table}[h!]
\vspace{-.5cm}
   \begin{center}
     \begin{tabular} {|p{2.15cm}|c|c|c|c|c|c|} \hline
 & $c_0$ & $c_1$ & $M_0$ & $\alpha$ & $\beta$ & Phase \\ 
\vspace{-.53cm}\centering{Background}
& $(GeV^{-1})$ & $(GeV^{1-\alpha})$ & $(GeV)$ &  & $(GeV)$ & (deg.)  \\
\hline
\centering{$F_{JS}$} & $84\pm 55$ & $900\pm 400$ & 
$1.65\pm 0.05$ & $0$ & $1.4\pm 0.2$ & $0$ (fixed) \\
\hline
       \end{tabular}
       \caption{Fit results with two Breit-Wigner and one Jacob-Slansky
     amplitudes.}
     \label{TAB:FIT1}
     \vfill
   \end{center}
 \end{table}
\begin{table}
   \begin{center}
     \begin{tabular} {|p{2.15cm}|p{3cm}|p{2.08cm}|c|c|} \hline
     & 
\centering{Mass} & \centering{Width} & $B_{ee}B_{3\pi^{+}3\pi^{-}}/M^2$ & Phase \\ 
\vspace{-.53cm}  
\centering{Resonances}  & \centering{(GeV/c$^2$)} & 
\centering{(MeV/c$^2$)} & 
\centering{(Yield/10 MeV)} & (deg.) \\ \hline
 \centering{$V_{0}$} & \centering{$1.910\pm 0.010 $} & 
\centering{$ 33\pm 13$} & $ 5\pm 2 $ & $84\pm 30$ \\
 \centering{$V_{1}$} & \centering{$1.650\pm 0.050 $} & 
\centering{$240\pm 80$} & $ 21\pm 4$ & $150 \pm 30$\\
\hline\hline
 & 
\centering{Mass} & \centering{Width} & $B_{ee}B_{3\pi^{+}3\pi^{-}}/M^2$ & Phase \\
\vspace{-.53cm}
\centering{Background} & \centering{(GeV/c$^2$)} & 
\centering{(MeV/c$^2$)} & 
\centering{(Yield/10 MeV)} & (deg.) \\ \hline
\centering{BW} &\centering{$2.250\pm 0.030 $}  &\centering{$830\pm 150$} 
& $ 24\pm 1$ & $ 0$ (fixed) \\ \hline
      \end{tabular}
     \caption{Fit results with three Breit-Wigner amplitudes.
   \label{TAB:FIT}} 
     \vfill
   \end{center}
 \end{table}
\par
\begin{figure}
 \centerline{
\epsfig{file=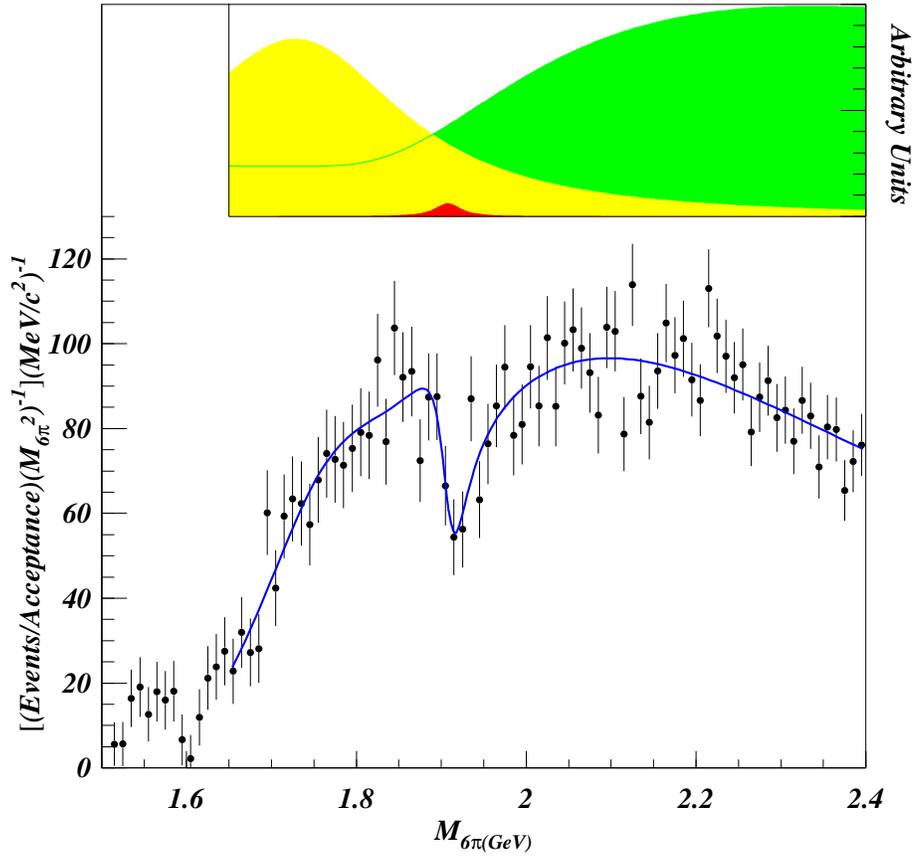,width=5.0in,height=5.0in}
 }
 \caption{ E687 $3\pi^+ 3\pi^- $ invariant mass distribution.
 Continuous line: fit with
  two resonances and Jacob-Slansky continuum
  (parameters in Tab.~\ref{TAB:FIT1}). 
 Inset: relative fraction of each amplitude without interference.   }
 \label{FIG:TWOBWONEJS}
\end{figure}
 \par
\section{Discussion and possible interpretations}
\begin{figure}
 \begin{center}
 \epsfig{file=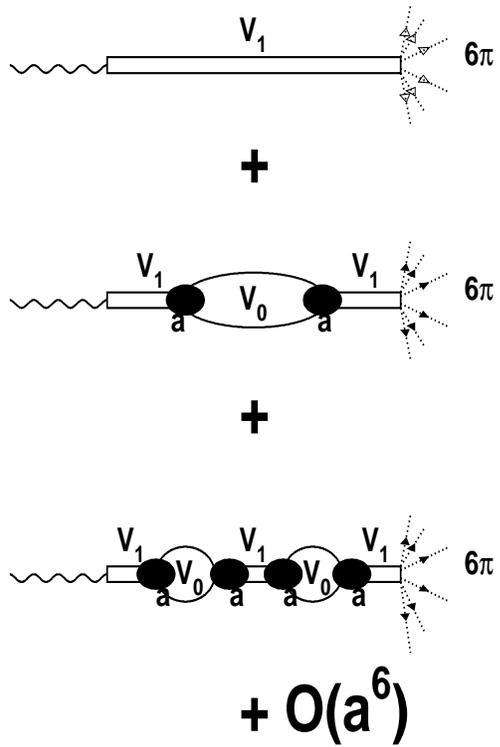,width=4.0in,height=5.0in}
 \end{center}
 \caption{ Diagram describing the $\epem$ annihilation via a
 $V_0, V_1$ interference term
 contribution. }
 \label{FIG:DIAGR}
\end{figure}

The narrow resonance $V_0$ pointed out by E687 has a small width and
a small production cross section, i.e., a small \ee partial width 
with respect to the broad prominent $\rho(1700)$ resonance.
In this situation a simple mixing mechanism can explain the dip structure
independently of the nature of $V_0$.  
In the extreme limit of full mixing, with the assumption 
that $V_0$ cannot couple directly
to the six-pion final state, the corresponding amplitude, as shown in
(Fig. \ref{FIG:DIAGR}), must  
include the propagator of a broad vector meson, say $V_1$, added to the
$V_0$ propagator times the coupling constant $a$ between $V_1$ and $V_0$, possibly
repeated:  
\bea
A & \propto & \frac{1}{M^2-M_1^2}
(1+a\frac{1}{M^2-M_0^2}a\frac{1}{M^2-M_1^2}+  \nonumber \\
  &         & + a\frac{1}{M^2-M_0^2}a\frac{1}{M^2-M_1^2}
a\frac{1}{M^2-M_0^2}a\frac{1}{M^2-M_1^2}+
\mathcal{O}(a^6)) \nonumber \\
 & \propto & \frac{M^2-M_{0}^2}{(M^2-M_{1}^2)(M^2-M_{0}^2) - a^2}. 
\eea
Here the six-pion invariant mass 
squared is $M^2$, the complex number $M$ stands for mass and width of any
$\rho$ recurrence $V_1$ nearby, $M_{0}$ 
is the
complex mass for the narrow resonance $V_{0}$.
This amplitude, with a zero at the unmixed $V_{0}$ mass pole $M_{0}$ in the
limit of negligible unmixed width, 
will produce a narrow dip at 
$\sqrt(s) \sim M_{0}$ in the 
cross section, which is consistent with what has been observed in the E687
analysis. 
This phenomenon was originally introduced at the time
 the toponium was expected on top of the $Z^0$ \cite{Gilman}. 
\par
It should be noted that the observation of $V_{0}$ strongly depends
on the interference mechanism. Fig. 2 reports in the inset what could
be expected if $V_{0}$ did not interfere with another broad resonance.
If this were the case, there would be no hope of detecting this resonance.
Therefore also in other channels the evidence strongly depends on the 
interference pattern, unless dynamical reasons make the coupling of $V_{0}$ 
to that specific channel very strong.
\par
We now discuss a physical interpretation of V$_0$.
This resonance cannot be interpreted as a glueball, a bound state of valence
gluons, 
because a glueball
is expected to be an isoscalar. Incidentally all the present
lattice calculations, in the quenched approximation, agree in predicting
the lightest isoscalar vector glueball at $\sim 4 ~GeV$.
\par
 This structure could be interpreted as multiquark or molecular state(s).
 These states should be narrow and clustered near the constituent total mass,
even if calculations have shown that they should not exist as resonances
\cite{Myh}, with 
some possible remarkable exception \cite{Isgur1}.
A particular case of multiquark states  is represented by $N \overline{N}$ bound
states and 
resonances, which  
should cluster at the $N \overline{N}$ threshold and $V_{0}$ is nearby.
Therefore a $N \overline{N}$ resonance has to be
considered and there is evidence of bumps in the $V_{0}$ mass region
\cite{franklin}. 
Recently new results from BES 
seem to indicate the presence of a structure in this energy region
\cite{Bai:2003sw}. 
However, OBELIX has looked for such a resonance in
$\overline{n} p \rightarrow 3 \pi^{+} 2 \pi^{-} \pi^{0}$
with a negative result \cite{Obelix}, 
so this
interpretation is unlikely, taking into account that the $N \overline{N}$ 
channel should be strongly coupled according to this interpretation.  The
possibility of having the narrow resonance out of the OBELIX narrow
 kinematical region of invariant masses, because of small 
 mass shifts between
 experiments, should also be kept in mind.
 \par
On the other hand, similar dips and narrow structures in other \ee annihilation
channels have been observed in this energy region and
the old argument supporting the existence of $N \overline{N}$ bound states
and resonances near threshold
is still very compelling
\cite{Dov}\cite{Sha}\cite{Jaf}\cite{Rich}\cite{Yan}.
In the case of vector $N \overline{N}$ states,
annihilation into $\epem$ may cross the threshold. 
Such bound states and resonances would
appear as a steep variation in the nucleon time-like form factors near 
threshold \cite{PS170} and a 
dip in the multihadronic cross section \cite{baldini1}. 
Indeed, a steep variation in the nucleon time-like form factors 
 and also a dip in the total $e^{+}e^{-}$ multihadronic cross section have been observed, 
in agreement with a narrow resonance at $\sim 1.87~ GeV$, just below the
$N \overline{N}$ threshold \cite{fenice}. However this baryonium candidate is hardly 
consistent with the E687 dip because of the $\sim 30~ MeV$ mass difference.
A cusp effect connected with the crossing of an unidentified threshold \cite{rosner}
could give rise to a steep downward step, followed however by
a slow rise, if any at all.
\par
 The $V_0$ could plausibly be interpreted as a hybrid, i.e. a $q \bar q g$
bound state.
Many theoretical approaches predict the existence of hybrid states
\cite{Felipe}.  
In the framework of the flux tube model 
\cite{Isgur}\cite{Barnes}, 
the hybrid new degree of freedom is identified in the excitation of
the color flux tube connecting the valence quarks. 
The flux tube model predicts nonstrange hybrids at $\sim 1.9 ~GeV/c^2$
and strange hybrids at $M \sim 2.1 ~GeV/c^2$. A similar prediction
has been obtained by lattice calculations 
\cite{UKQCD1}\cite{Bernard}\cite{Lacock}.
Small, but not vanishing, e.m. widths characterize hybrids, since the
gluon does not couple to the photon.
The way the string breaks forbids decay into two identical mesons and 
imposes spin and parity of the decay products \cite{Page}. 
Because of these selection rules in two-body decay, high multiplicity
channels should be preferred and a relatively small width foreseen.
Narrow hybrids are predicted,
in particular a vector isoscalar hybrid, a few MeV wide, still at $\sim 1.9~ GeV$ 
\cite{Page}.
\par
On the other hand, it is not unanimously agreed that valence gluons  exist at all.
In the $1/N_{color}$ expansion there is no suppression of gluon
creation and it has been claimed there is 
no reason to expect valence gluons \cite{Chan}.
It has also been argued that in classical field theory, pure gauge bound states
are not likely to exist: in fact, in analogy with electric charges, internal
directions somewhere become antiparallel, whereas continuity requires close fields
pointing in the same direction \cite{Coleman}.
On the contrary, valence gluons are naturally foreseen if confinement is 
properly described by the bag model \cite{Chan}.
\par
Diffractive photoproduction was recently pointed out as a powerful tool to
search for hybrids \cite{Szczepaniak:2001qz}. Future searches both at high
and low energy should particularly address both the confirmation of the E687
effect as well as the search for the $1^{--}$ isoscalar partner of the
E687 state, 
i.e. any effect in the invariant mass distribution of an odd-number of pions.
\par
\section{Conclusions}
We have investigated the nature of the dip structure observed by E687 in
diffractive photoproduction. A coherent fit of two BW resonances plus
an $F_{JS}(M)$ amplitude of the E687 data is consistent with a narrow 
resonance strongly interfering with known vector mesons, such as $\rho(1700)$. 
We have pointed out that in this scenario such a resonance has to appear as a 
dip in the mass spectrum. An interpretation of the $V_0$ as a $1^{--}$, isovector
hybrid is in agreement with expected mass, width, and decay mode.
A $N \overline{N}$ resonance is unlikely according to OBELIX, which has looked 
for such a resonance in $\overline{n} p \rightarrow 3 \pi^{+} 2 \pi^{-} \pi^{0}$
with a negative result.

\end{document}